\documentclass{article}

\usepackage{times}
\usepackage{stmaryrd}
\usepackage{epsfig}
\usepackage{amsfonts}
\usepackage{amsmath}
\usepackage{latexsym}
\usepackage{amssymb}
\usepackage[mathcal]{euscript}
\usepackage{eurosym}
\usepackage{rotating}

\newcommand{\oar}{\rhd}
\newcommand{\true}{\textrm{\textbf{true}}}
\newcommand{\false}{\textrm{\textbf{false}}}
\newcommand{\va}{\textrm{\textbf{var}}}
\newcommand{\lr}{\longrightarrow}

\newcommand{\ket}[1]{| #1 \rangle}
\newcommand{\bra}[1]{\langle #1 |}

\newcommand{\trace}{\textrm{Tr}}

\newcommand{\europ}{\textup{\emph{\geneuro}}}

\newtheorem{Def}{Definition}

\newtheorem{Proposition}{Proposition}

\begin{document}

\title{Linear-algebraic $\lambda$-calculus}

\author{Pablo Arrighi\thanks{
Institut Gaspard Monge,
5 Bd Descartes, Champs-sur-Marne,
77574 Marne-la-Vall\'ee Cedex 2, France,
{\tt arrighi@univ-mlv.fr.}}
\and
Gilles Dowek\thanks{
\'Ecole polytechnique and INRIA,
LIX, \'Ecole polytechnique,
91128 Palaiseau Cedex, France,
{\tt Gilles.Dowek@polytechnique.fr.}}}
\date{\vspace{-.5in}}

\maketitle \thispagestyle{empty}

\begin{abstract}
With a view towards models of quantum computation and/or the
interpretation of linear logic, we define a functional language
where all functions are linear operators by construction. A small
step operational semantic (and hence an interpreter/simulator) is
provided for this language in the form of a term rewrite system.
The \emph{linear-algebraic $\lambda$-calculus} hereby constructed
is linear in a different (yet related) sense to that, say, of the
linear $\lambda$-calculus. These various notions of linearity are
discussed in the context of quantum programming languages.
\end{abstract}

\section{Introduction}

Quantum computation lacks a convenient model of computation. To
this day its algorithms are expressed in terms of quantum
circuits, but their descriptions always seem astonishingly remote
from the task they do accomplish \cite{Deutsch}. Moreover
universality is only provided via the notion of uniform family of
circuits \cite{Yao}. Quantum Turing machines solve this latter
point, yet they are even less suitable as a programming language
\cite{Bernstein}. Another approach is to enclose quantum circuits
within a classical imperative-style control structure
\cite{Selinger} --- but we wish to avoid this duality, in an attempt
to bring programs closer to their specifications. Functional-style
control structure, on the other hand, seem to merge with quantum
evolution descriptions in a unifying manner. With a view towards
models of quantum computation, we describe a functional language
for expressing linear operators, and linear operators only.

We are careful, however, not to bury our presentation of this
language of linear operators within too many quantum
computation-specific considerations. The aim is to reach an
audience of logicians also, as we suspect a strong connection with
issues of computational interpretations of linear logic.

We provide a semantic for the language in the form of a term
rewrite system \cite{Dershowitz}. These consist in a finite set of
rules $l\lr r$, each interpreted as follows: \emph{``Any term $t$
containing a subterm $\sigma l$ in position $p$ (i.e. $t=t[\sigma
l]_p$) should be rewritten into a term $t'$ containing $\sigma r$
in position $p$, with all the rest unchanged (i.e. $t'=t[\sigma
r]_p$)''.} Here $\sigma$ denotes a variable substitution.  The
minimalist interpretation of the rules makes term rewrite systems
(TRS) extremely suitable for describing the behavior of a computer
languages unambiguously --- so long as the order in which the
reductions occur does not matter to the end result (a property
named confluence). Moreover, because $l\lr r$ may be seen as an
oriented version of equation $l=r$, the TRS provides both an
operational semantic (an interpreter/simulator for the language)
and an axiomatic semantic (an equational theory in which to prove
properties about the language).

We begin with a simple language for vectors containing constants
for base vectors, addition and product by a scalar. On terms of
this language we define a rewrite system reducing any term
expressing a vector to a linear combination of base vectors
\cite{arrighidowek}. We have also proposed in \cite{arrighidowek}
an extension to a language containing a tensorial product
operation (Section \ref{sectionwrla}).

Such a language must rely on a language and rewrite system for
scalars. This raises the problem of the conditional rewriting
required for division, which we can circumvent, basing quantum
computation upon the ring of diadic floats together with
$\frac{1}{\sqrt{2}}$ and imaginary number $i$ (Section
\ref{sectionfield}). More generally, it should be said that a
language of linear operators does not need division.\\

Modern days functional languages such as Caml, Haskell etc. are
based upon two basic evaluation mechanisms: \emph{matching}, which
provides conditional branching by inspection of values;
and some avatar of the \emph{$\lambda$-calculus}.
The first mechanism is obtained as we extend the term rewrite
system to handle linear maps --- themselves denoted as
superpositions of bipartite states, e.g.
\begin{equation*}
\left( \true \oar \false + \false \oar \true \right)* \false \lr^*
\true.
\end{equation*}
Applications are therefore analogous to contractions in tensorial
calculus: this approach offers an elegant paradigm to represent
quantum operations as quantum states (Section
\ref{sectionmatching}).\\
The second mechanism is obtained through an implementation of
$\lambda$-terms via de Bruijn indices, a scheme whereby variables
are encoded as integers referring to their binders, e.g.
\begin{equation*}
\lambda x.(\lambda y. (x \otimes y))\textrm{  is encoded as  }
L(L(\va(1) \otimes \va(0))).
\end{equation*}
The question of the interpretation of terms such as $\lambda
x.(x \otimes x)$ is lengthily addressed as we draw a
distinction between \emph{cloning} and \emph{copying}. The
semantic of our calculus forbids only the former, non-linear
operation, by enforcing a higher priority of the addition's
distributivity over substitution (Section \ref{sectionlambda}).
This is followed by a short example taken from our implementation (Section \ref{example}).

\emph{Erasure} on the other hand remains allowed in our calculus,
because we do not restrict ourselves to unitary operations. Whilst
we discuss possible well-formedness conditions to implement this
restriction (a crucial one for quantum computation), the claim
here is to have provided a ``\emph{linear}'' $\lambda$-calculus,
in the sense of \emph{linear algebra}. We discuss the various
notions of ``linearity'' used in quantum programming languages, such
as the one by Van Tonder \cite{Van Tonder1} (Section
\ref{sectiondiscussion}).

\section{Vectorial spaces}
\label{sectionwrla}

We seek to represent quantum programs, their input vectors, their
output vectors and their applications as terms of a first-order
language.  Moreover we seek to provide rules such that the term
formed by the application of a quantum program onto its input
vector should reduce to its output vector. Several terms may be
used to express one output vector, as a consequence we must ensure
that these all reduce to one unique, \emph{normal form}, upon
which there is nothing more to compute. The most natural normal
form to aim for is that of a linear combination of the base
vectors, i.e. the computation finishes once we have the
coordinates of the output vector.

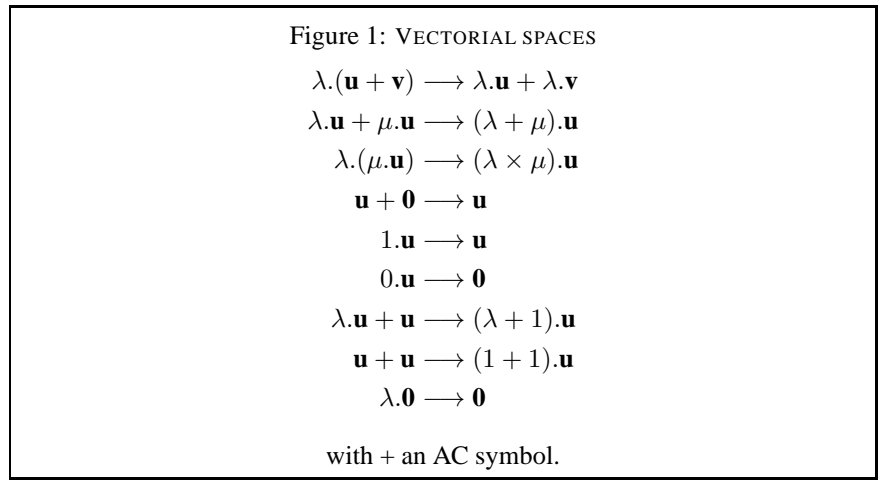
\begin{figure}[!ht]
\center{\fbox{\begin{minipage}[t]{11.3cm}\vspace{-2mm}
\caption{\label{rulesvecsp}\small\textsc{Vectorial spaces}}\vspace{-6mm}
\begin{align*}
\lambda . (\textbf{u} + \textbf{v}) &\lr \lambda . \textbf{u} +
\lambda . \textbf{v}\\
\lambda . \textbf{u} + \mu . \textbf{u} &\lr (\lambda+\mu) .
\textbf{u}\\
\lambda . (\mu . \textbf{u}) &\lr (\lambda\times\mu) . \textbf{u}\\
\textbf{u} + \textbf{0} &\lr \textbf{u}\\
1 . \textbf{u} &\lr
\textbf{u}\\
0 . \textbf{u} &\lr \textbf{0}\\
\lambda . \textbf{u} + \textbf{u} &\lr (\lambda+1) . \textbf{u}\\
\textbf{u} + \textbf{u} &\lr (1+1) . \textbf{u}\\
\lambda . \textbf{0} &\lr \textbf{0}
\end{align*}
\center{with + an AC symbol.}
\end{minipage}}}
\end{figure}

We start with the language of vectorial spaces, i.e. a two-sorted
language $\mathcal{L}$ having sort $K$ for scalars and sort $E$
for vectors --- together with: two constants $0$ and $1$ of sort
$K$; a constant $\mathbf{0}$ of sort $E$; two binary symbols $+$
and $\times$ of rank $\langle K,K,K \rangle$; a binary symbol $+$
(also) of rank $\langle E,E,E \rangle$; and a binary symbol $.$ of
rank $\langle K,E,E \rangle$. In \cite{arrighidowek} we described
a term rewrite system reducing any term expressing a vector into a
linear combination of base vectors. The term rewrite system
develops
$$4 . (\textbf{false} + \textbf{true}) \lr 4 . \textbf{false} + 4
. \textbf{true}$$
but factorizes
$$ 4 . \textbf{false} + 6 . \textbf{false} \lr (4 + 6) . \textbf{false}.$$
according to the rules in figure \ref{rulesvecsp}. Such a TRS
arises as we orient six of the eight equations axiomatizing
vectorial spaces. Only those two axioms corresponding to
associativity and commutativity of vector addition are left aside, 
because we use rewriting modulo AC(+). Moreover we need to add
three more rules for confluence.

But these rewrite rules do not take into account computation on
scalars. The latter must be added by mixing in another rewrite
system $S$, rewriting scalar to a normal form.

\begin{Def}(Scalar rewrite system)
A {\em scalar rewrite system} is a rewrite system on a language
containing at least the symbols $+$, $\times$, $0$ and $1$, such
that:

\begin{itemize}
\item $S$ is terminating and ground confluent,

\item for all closed terms
$\lambda$, $\mu$ and $\nu$, the pair of terms
\begin{itemize}
\item $0 + \lambda$ and $\lambda$,
\item $0 \times \lambda$ and $0$,
\item $1 \times \lambda$ and $\lambda$,
\item $\lambda \times (\mu + \nu)$ and $(\lambda \times \mu) +
(\lambda \times \nu)$,
\item $(\lambda + \mu) + \nu$ and $\lambda + (\mu + \nu)$,
\item $\lambda + \mu$ and $\mu + \lambda$,
\item $(\lambda \times \mu) \times \nu$ and $\lambda \times (\mu \times \nu)$,
\item $\lambda \times \mu$ and $\mu \times \lambda$
\end{itemize}
have the same normal forms,

\item $0$ and $1$ are normal terms.
\end{itemize}
\end{Def}

The following propositions can be found in \cite{arrighidowek}.

\begin{Proposition}\label{proptermconf}
For any scalar rewrite system $S$,
the rewrite system $R \cup S$ is terminating and ground confluent.
\end{Proposition}

\begin{Proposition}\label{propnorm}
If $t$ is a normal close term whose constants are amongst
${\bf x}_{1}, ..., {\bf x}_{n}$. The term ${\bf t}$ is ${\bf 0}$
or it has the form $\lambda_{1} {\bf x}_{i_{1}} + ... \lambda_{k}
{\bf x}_{i_{k}} + {\bf x}_{i_{k+1}} + {\bf x}_{i_{k+l}}$ where the
indices $i_{1}, ..., i_{k+l}$ are distinct and the $\lambda_{k}$'s
are neither $0$ nor $1$.
\end{Proposition}

Note that the algorithm defined by $R$ is relatively common in
computing, for presenting any vector as a linear combination of
base vectors. But it does in fact define vectorial spaces, as any
mathematical structure validating the algorithm. In this sense we
have provided a computational definition of vectorial spaces.

Furthermore note that the support for tensor products is easily
added into the TRS, through the six rules given in figure
\ref{rulestensors}. Proposition \ref{proptermconf} remains true
when $R$ is extended with those six additional rules, whilst
proposition \ref{propnorm} now yields normal forms for terms in $E
\otimes E$ of the form $\mathbf{0}$ or
$$\lambda_{1} {\bf x}_{i_{1}} \otimes {\bf y}_{j_{1}} +\ldots+
\lambda_{k} {\bf x}_{i_{k}} \otimes {\bf y}_{j_{k}} + {\bf
x}_{i_{k+1}} \otimes {\bf y}_{j_{k+1}} +\ldots +{\bf
x}_{i_{k+l}} \otimes {\bf y}_{j_{k+l}},$$ where the pairs of
indices $\langle i_1, j_1 \rangle,\ldots, \langle i_{k+l}, j_{k+l} \rangle$ are distinct and
the $\lambda_k$'s are neither $0$ nor $1$ \cite{arrighidowek}.

\begin{figure}[!h]
\center{\fbox{\begin{minipage}[t]{11.3cm}\vspace{-2mm}
\caption{\label{rulestensors}\small\textsc{Vectorial spaces:
tensors}}\vspace{-6mm}
\begin{align*}
({\bf u} + {\bf v}) \otimes {\bf w} &\lr {\bf u} \otimes {\bf w} +
{\bf v} \otimes {\bf w}\\
(\lambda .{\bf u}) \otimes {\bf v} &\lr \lambda .({\bf u} \otimes
{\bf v})\\
{\bf u}  \otimes ({\bf v} + {\bf w}) &\lr
{\bf u} \otimes {\bf v} + {\bf u} \otimes {\bf w}\\
{\bf u} \otimes (\lambda .{\bf v}) &\lr
\lambda .({\bf u} \otimes {\bf v})\\
{\bf 0} \otimes {\bf u} &\lr {\bf 0}\\
{\bf u} \otimes {\bf 0} &\lr {\bf 0}
\end{align*}
\end{minipage}}}
\end{figure}

\section{The field of quantum computing}
\label{sectionfield}

Fields are not easily implemented as term rewrite systems, because
of the conditional rewriting required for the division by zero. In
the previous section such problems were avoided by simply assuming
a TRS for scalars having a certain number of properties, but if
the objective is to lay the ground for formal quantum programming
languages, then we must provide such a TRS. The present section
briefly outlines how this is achieved.

\subsection{Background}

We seek to model quantum computation as a formal rewrite system
upon a finite set of symbols. Since the complex numbers are
uncountable, we must therefore depart from using the whole of
$\mathbb{C}$ as the field $\mathbb{K}$ of our vectorial space. Such
considerations are commonplace in computation theory, and were
successfully addressed with the provision of the first rigorous
definition of a quantum Turing machine \cite{Bernstein}. In short
the quantum Turing machines are brought as an extension of
probabilistic Turing machines
\begin{align*}
\langle &\!Q\!:\textrm{head states},\,\Sigma\!: \textrm{alphabet},\\
&\,\delta\!: \textrm{transition function}, \,q_o,q_f\!:
\textrm{start,end state}\!\!\rangle
\end{align*}
whose transition functions are no longer valued over the
efficiently computable positive reals (probabilities)
\begin{equation*}
\delta: Q\times\Sigma\longrightarrow
(Q\times\Sigma\times\{Left,\,Right\}\rightarrow
\tilde{\mathbb{R}}^{+})
\end{equation*}
but over the efficiently computable complex numbers (amplitudes)
\begin{equation*}
\delta: Q\times\Sigma\longrightarrow
(Q\times\Sigma\times\{Left,\,Right\}\rightarrow
\tilde{\mathbb{C}}).
\end{equation*}
In both cases $\delta$ is constrained to be a \emph{unit} function
(probabilities/squared modulus summing to one), and for the
quantum Turing machine $\delta$ is additionally required to induce
a \emph{unitary} global evolution. A well-known result of
complexity theory is that probabilistic Turing machines remain as
powerful when the transition function $\delta$ is further
restricted to take values in the set $\{0,\frac{1}{2},1\}$. The
result in \cite{Bernstein} is analogous: quantum Turing machines
remain as powerful when the transition function $\delta$ is
further restricted to take values in the set
$\{-1,-\frac{1}{\sqrt{2}},0,\frac{1}{\sqrt{2}},1\}$. Later it was
shown in \cite{Adleman}, and independently in \cite{Solovay2} that
no irrational number is necessary, i.e. $\delta$ may be restricted
to take values in the set
$\{-1,-\frac{8}{5},-\frac{3}{5},0,\frac{3}{5},\frac{8}{5},1\}$
without loss of power for the quantum Turing machine.

In the circuit model of quantum computation the emphasis was
placed on the ability to \emph{approximate} any unitary transform
from a finite set of gates. This line of research (cf.
\cite{Solovay1}\cite{Kitaev} to cite a few) has so far culminated
with \cite{Boykin}, where the following set
\begin{align}
CNOT&=\left(\begin{array}{cccc}
1    &0    &0   &0\\
0    &1    &0   &0\\
0    &0    &0   &1\\
0    &0    &1   &0 \end{array}\right)\label{basicgates}\\
H&=\left(\begin{array}{cc}
\frac{1}{\sqrt{2}}     &\frac{1}{\sqrt{2}} \\
\frac{1}{\sqrt{2}}     &-\frac{1}{\sqrt{2}}
\end{array}\right)\quad
P=\left(\begin{array}{cc}
1     &0 \\
0     &e^{i\pi/4}
\end{array}\right)\nonumber
\end{align}
was proven to be universal in the above strict sense. A weaker
requirement for a set of gates is the ability to \emph{simulate}
any unitary transform, a notion which is also referred to as
\emph{encoded universality} --- since a computation on $n$ qubits
may for instance be represented as a computation on $n+1$ ``real
bits'', through a simple mapping. A recent paper shows that the
gate
\begin{align*}
G=\left(\begin{array}{cccc}
1    &0    &0   &0\\
0    &1    &0   &0\\
0    &0    &a   &-\!b\\
0    &0    &b   &a \end{array}\right),
\end{align*}
with either $a=b=\frac{1}{\sqrt{2}}$, or $a=\frac{3}{5}$ and
$b=\frac{8}{5}$, has this property \cite{Rudolph}. Do appreciate
how the result falls into line with those regarding the quantum
Turing machine.
\begin{Def}\label{defcompscals}
We call {\em computational scalars}, and denote $\tilde{\mathbb{K}}$ the
ring formed by the additive and multiplicative closure of the
complex numbers $\{-1,1,\frac{1}{\sqrt{2}},i\}$.
\end{Def}
Once we have shown that the computational scalars arithmetics can
be performed by a TRS, it will be sufficient to express the basic
gates (\ref{basicgates}) in our formalism to immediately obtain
the more traditional notion of quantum computation universality.
Hence our choice.

\subsection{Rules}
We begin by implementing natural numbers and unsigned binary
numbers. That such TRS can be made ground confluent and
terminating are now well-established results
\cite{Cohen}\cite{Walters}. This places us in a position to build
up diadic floats out of a sign, an unsigned binary number and an
exponent, e.g. $\textrm{fl}(\textrm{neg},1,S(\textrm{zeron}))$ is
to stand for $-\frac{1}{2}$, as exemplified in figure
\ref{rulesfloats}.

\begin{figure}[!h]
\center{\fbox{\begin{minipage}[t]{11.3cm}\vspace{-2mm}
\caption{\label{rulesfloats}\small\textsc{Diadic
floats}}\vspace{-6mm}
\begin{align*}
\textrm{fl}(s,n::0,S(p)) &\lr \textrm{fl}(s,n,p)\\
\textrm{fl}(\textrm{neg},0,p) &\lr
\textrm{fl}(\textrm{pos},0,p)\\
\textrm{fl}(s,0,S(p)) &\lr \textrm{fl}(s,0,\textrm{zeron})\\
\vdots\\
\textrm{fl}(\textrm{pos},m_1,e_1) \textrm{ timesf }
\textrm{fl}(\textrm{neg},m_2,e_2) &\lr \textrm{fl}(\textrm{neg},
m_1\textrm{ timesb
}m_2,\textrm{addn}(e_1, e_2))\\
\textrm{fl}(\textrm{neg},m_1,e_1) \textrm{ timesf }
\textrm{fl}(\textrm{pos},m_2,e_2) &\lr
\textrm{fl}(\textrm{neg},m_1\textrm{ timesb }m_2,\textrm{addn}(e_1, e_2))\\
\vdots
\end{align*}
\end{minipage}}}
\end{figure}

Reached this point it suffices to notice that
$\tilde{\mathbb{K}}$, i.e. diadic floats together with imaginary
number $i$ and real number $\frac{1}{\sqrt{2}}$, can be viewed as
a four-dimensional module upon diadic floats. Indeed any such
number could be represented as a linear combination of the form:
\begin{equation*}
\alpha . \mathbf{1} + \beta . \mathbf{\frac{1}{\sqrt{2}}} + \gamma
. \mathbf{i} + \delta . \mathbf{\frac{i}{\sqrt{2}}}.
\end{equation*}
As a consequence we can reuse the results of section
\ref{sectionwrla} to implement computational scalars and their
additions. Computational scalars multiplication then needs to be
defined, we do so modulo AC in figure \ref{rulesscals}. Notice that we
overload the symbol $\times$ for multiplication of diadic floats and 
for multiplication of computational scalars.

\begin{figure}[!h]
\center{\fbox{\begin{minipage}[t]{11.3cm}\vspace{-2mm}
\caption{\label{rulesscals}\small\textsc{Scalar
multiplication}}\vspace{-6mm}
\begin{align*}
\mathbf{1} \times \mathbf{v} &\lr \mathbf{v}\\
\mathbf{\frac{1}{\sqrt{2}}}  \times  \mathbf{\frac{1}{\sqrt{2}}} &\lr
\textrm{fl}(\textrm{pos},1,S(\textrm{zeron})).\mathbf{1}\\
\mathbf{\frac{1}{\sqrt{2}}} \times  \mathbf{i} &\lr \mathbf{\frac{i}{\sqrt{2}}}\\
\mathbf{\frac{1}{\sqrt{2}}} \times  \mathbf{\frac{i}{\sqrt{2}}} &\lr
\textrm{fl}(\textrm{pos},1,S(\textrm{zeron})).\mathbf{i}\\
\mathbf{i}  \times  \mathbf{i} &\lr \textrm{fl}(\textrm{neg},1,\textrm{zeron}).\mathbf{1}\\
\mathbf{i}  \times  \mathbf{\frac{i}{\sqrt{2}}} &\lr \textrm{fl}(\textrm{neg},1,\textrm{zeron}).\mathbf{\frac{1}{\sqrt{2}}}\\
\mathbf{\frac{i}{\sqrt{2}}} \times \mathbf{\frac{i}{\sqrt{2}}} &\lr
\textrm{fl}(\textrm{neg},1,S(\textrm{zeron})).\mathbf{1}
\vdots\\
(\lambda.\mathbf{u})  \times  \mathbf{v} &\lr
\lambda.(\mathbf{u} \times \mathbf{v})\\
(\mathbf{t}+\mathbf{u}) \times \mathbf{v} &\lr \mathbf{t} \times \mathbf{v} +
\mathbf{u} \times \mathbf{v}
\end{align*}
\center{with  $\times$  an AC symbol.}
\end{minipage}}}
\end{figure}

We conjecture that this TRS is ground confluent and terminating, but
have not yet a formal proof for this assertion.

Notice we have never defined a division operation. This is because
only the ring properties of these numbers are required for
expressing linear operations: we place ourselves upon a ``module''
rather that a full vectorial space.

\section{Matching construct}
\label{sectionmatching}

We now turn to the definition of the matching constructs in our
language. As we shall see, these constructs are nothing else than 
a reformulation of the rules for the tensor product.

\subsection{Notations}

Your typical functional language (\texttt{Haskell},
\texttt{ML}\ldots) will always have ``matching'' constructs (for
branching). For instance, here is a piece of
\texttt{Caml}:\\
{\tt let rec not b = match b with\\
| false -> true\\
| true -> false ;;}\\
We wish to provide such constructs in our linear-algebraic
calculus. Strangely enough these matching constructs are very
close to the tensorial product constructs.

Mathematicians and physicist in this field would
write linear maps instead:
$\textrm{NOT}=\ket{\textrm{true}}\bra{\textrm{false}} +
\ket{\textrm{false}}\bra{\textrm{true}}$. However here the
$\bra{\textrm{false}}$ and $\bra{\textrm{true}}$ may be viewed as
patterns, waiting to be compared to the input vector through a
scalar product. Thus we choose to reconcile both worlds and write:
$$\textrm{NOT}=\textbf{false}\rhd\textbf{true}+\textbf{true}\rhd\textbf{false}.$$
An expression $(\textbf{t}\rhd\textbf{u})$ applied to a vector
$\textbf{v}$ will then reduce into $(\textbf{t}\bullet
\textbf{v}).\textbf{u}$, with $\bullet$ the scalar product. In this sense
$(\textbf{t}\rhd\textbf{u})*\textbf{v}$ does return $\textbf{u}$
\emph{in so far as} $\textbf{t}$ overlaps with $\textbf{v}$. More
formal justifications, and formal rewrite rules follow in the next
two subsections. For now we give the reduction steps involved in
the application of the phase gate $P$ upon the vector
\textbf{true}, as a motivating example for these rules:
\begin{align*}
&\left((\textbf{false} \rhd \textbf{false})+ \textbf{true} \rhd
(\frac{1}{\sqrt{2}}+i\frac{1}{\sqrt{2}}).\textbf{true}\right)
* \textbf{true}\\
\lr^*&\;\; (\textbf{false} \rhd \textbf{false}) * \textbf{true} +
\big(\textbf{true}\rhd(\frac{1}{\sqrt{2}}+i\frac{1}{\sqrt{2}}).\textbf{true}\big)
* \textbf{true}\\
\lr^*&\;\; (\textbf{false}\bullet\textbf{true}).\textbf{false} +
(\textbf{true}\bullet\textbf{true}).\big((\frac{1}{\sqrt{2}}+i\frac{1}{\sqrt{2}}).\textbf{true}\big)\\
\lr^*&\;\;
0.\textbf{false}+(\frac{1}{\sqrt{2}}+i\frac{1}{\sqrt{2}}).\textbf{true}\\
\lr^*&\;\; (\frac{1}{\sqrt{2}}+i\frac{1}{\sqrt{2}}).\textbf{true}.
\end{align*}
All of the three gates forming a universal set for quantum
computation are trivially expressed as terms in this notation:
\begin{align*}
CNOT =&\; (\textbf{false} \otimes \textbf{false}) \rhd (\textbf{false} \otimes \textbf{false})\\
&+ (\textbf{false} \otimes \textbf{true}) \rhd (\textbf{false} \otimes \textbf{true})\\
&+ (\textbf{true} \otimes \textbf{false}) \rhd (\textbf{true} \otimes \textbf{true})\\
&+ (\textbf{true} \otimes \textbf{true}) \rhd (\textbf{true} \otimes \textbf{false})\\
H=&\; \left(\textbf{false} \rhd \frac{1}{\sqrt{2}}.(\textbf{false}+\textbf{true})\right)+\left(\textbf{true} \rhd \frac{1}{\sqrt{2}}.(\textbf{false}-\textbf{true})\right)\\
P=&\; (\textbf{false} \rhd \textbf{false})+\left(\textbf{true}
\rhd
(\frac{1}{\sqrt{2}}+i\frac{1}{\sqrt{2}}).\textbf{true}\right).
\end{align*}

\subsection{Rules}

Since $\rhd$ is just another type of tensor product, bilinearity
applies (see figure \ref{rulesmatchingbilinearity}. Notice the
conjugation of the $\lambda$ scalar, denoted $\overline{\lambda}$,
easily implemented in the TRS).
\begin{figure}[!hp]
\center{\fbox{\begin{minipage}[t]{11.3cm}\vspace{-2mm}
\caption{\label{rulesmatchingbilinearity}\small\textsc{Matching
operators bilinearity}}\vspace{-6mm}
\begin{align*}
(\mathbf{t}+\mathbf{u})\rhd\mathbf{v} &\lr
\mathbf{t}\rhd\mathbf{v} + \mathbf{u}\rhd\mathbf{v}\\
\mathbf{t}\rhd(\mathbf{v}+\mathbf{w}) &\lr
\mathbf{t}\rhd\mathbf{v} + \mathbf{t}\rhd\mathbf{w}\\
(\lambda.\mathbf{u})\rhd\mathbf{v} &\lr
\overline{\lambda}.(\mathbf{u}\rhd\mathbf{v})\\
\mathbf{u}\rhd(\mu.\mathbf{v}) &\lr
\mu.(\mathbf{u}\rhd\mathbf{v})\\
\mathbf{0} \rhd \mathbf{u} &\lr \mathbf{0}\\
\mathbf{u} \rhd \mathbf{0} &\lr \mathbf{0}\\
({\bf u} + {\bf v}) * {\bf w} &\lr {\bf u} * {\bf w} +
{\bf v} * {\bf w}\\
(\lambda .{\bf u}) * {\bf v} &\lr \lambda .({\bf u} *
{\bf v})\\
{\bf u}  * ({\bf v} + {\bf w}) &\lr
{\bf u} * {\bf v} + {\bf u} * {\bf w}\\
{\bf u} * (\lambda .{\bf v}) &\lr
\lambda .({\bf u} * {\bf v})\\
{\bf 0} * {\bf u} &\lr {\bf 0}\\
{\bf u} * {\bf 0} &\lr {\bf 0}\\
(\mathbf{t}+\mathbf{u})\bullet\mathbf{v} &\lr
\mathbf{t}\bullet\mathbf{v} + \mathbf{u}\bullet\mathbf{v}\\
\mathbf{t}\bullet(\mathbf{v}+\mathbf{w}) &\lr
\mathbf{t}\bullet\mathbf{v} + \mathbf{t}\bullet\mathbf{w}\\
(\lambda.\mathbf{u})\bullet\mathbf{v} &\lr
\overline{\lambda}.(\mathbf{u}\bullet\mathbf{v})\\
\mathbf{u}\bullet(\mu.\mathbf{v}) &\lr
\mu.(\mathbf{u}\bullet\mathbf{v})\\
\mathbf{0} \bullet \mathbf{u} &\lr \mathbf{0}\\
\mathbf{u} \bullet \mathbf{0} &\lr \mathbf{0}
\end{align*}
\end{minipage}}}
\end{figure}
Other than its left-hand-side antilinearity, the particularity of
$\rhd$ is the reduction it induces when placed left of an
application symbol *, as described in figure
\ref{rulesmatchingtoscalar}. For definiteness, we may also add the
rules in figure \ref{rulesmatchingtoscalar2}. 
\begin{figure}[t]
\center{\fbox{\begin{minipage}[t]{11.3cm}\vspace{-2mm}
\caption{\label{rulesmatchingtoscalar}\small\textsc{Matching
operator and the scalar product }}\vspace{-6mm}
\begin{align*}
(\mathbf{t} \rhd \mathbf{u}) *\mathbf{v} &\lr (\mathbf{t} \bullet\mathbf{v}).\mathbf{u}\\
(\mathbf{t} \otimes \mathbf{u}) \bullet (\mathbf{v} \otimes
\mathbf{w}) &\lr (\mathbf{t} \bullet\mathbf{v}) \times (\mathbf{u}
\bullet \mathbf{w})\\
(\mathbf{t} \rhd\mathbf{u}) \bullet (\mathbf{v} \rhd \mathbf{w})
&\lr \overline{(\mathbf{t} \bullet\mathbf{v})} \times (u
\bullet \mathbf{w})\\
\textbf{true} \bullet \textbf{true} &\lr 1\\
\textbf{true} \bullet \textbf{false} &\lr 0\\
\textbf{false} \bullet \textbf{true} &\lr 0\\
\textbf{false} \bullet \textbf{false} &\lr 1
\end{align*}
\end{minipage}}}
\end{figure}

\begin{figure}[t]
\center{\fbox{\begin{minipage}[t]{11.3cm}\vspace{-2mm}
\caption{\label{rulesmatchingtoscalar2}\small\textsc{
Orthogonality rules}}\vspace{-6mm}
\begin{align*}
(\mathbf{t} \otimes \mathbf{u}) \bullet (\mathbf{v} \rhd \mathbf{w}) &\lr 0\\
(\mathbf{t} \otimes \mathbf{u}) \bullet \textbf{true} &\lr 0\\
(\mathbf{t} \otimes \mathbf{u}) \bullet \textbf{false} &\lr 0\\
(\mathbf{t} \rhd\mathbf{u}) \bullet (\mathbf{v} \otimes \mathbf{w}) &\lr 0\\
(\mathbf{t} \rhd\mathbf{u}) \bullet \textbf{true} &\lr 0\\
(\mathbf{t} \rhd\mathbf{u}) \bullet \textbf{false} &\lr 0\\
\textbf{true} \bullet (\mathbf{v} \otimes \mathbf{w}) &\lr 0\\
\textbf{true} \bullet (\mathbf{v} \rhd \mathbf{w}) &\lr 0\\
\textbf{false} \bullet (\mathbf{v} \otimes \mathbf{w}) &\lr 0\\
\textbf{false} \bullet (\mathbf{v} \rhd \mathbf{w}) &\lr 0
\end{align*}
\end{minipage}}}
\end{figure}
Notice that for now programming language does not use any
variables. All functions are defined by adding elementary functions
mapping base vectors to base vectors. And all functions are linear by
constructions. Moreover until Section \ref{sectiondiscussion} we do not worry about
normalization and unitarity conditions. For now the  the $\rhd$ notation
provides exactly what is needed: linear operations can be encoded
as sums of tensor states describing which vector is associated to which \cite{Arrighi1}\cite{Choi}.

\section{Lambda calculus construct}
\label{sectionlambda}

We now turn to the last ingredient of our language: the
lambda-calculus constructs. In this case we will need to introduce
bound variables. These bound variables will be handled through the
use of de Bruijn indices. We shall also be more careful, as
lambda-calculus usually allows to express non linear functions,
whilst we have to restrict ourselves to linear ones.

\subsection{Cloning}
\label{subsectioncloning} As we seek to provide a language of
linear operator, we must pay attention to the fact that
duplication of a vector can lead to nonlinear evolutions. For
instance, whatever formalism we choose for quantum theory (vectors
or density matrices), quantum operations act linearly upon their
input states. This, in turn, implies that quantum states cannot be
\emph{cloned}. Indeed such an evolution acts upon a qubit as
follows:
\begin{align*}
(\alpha\ket{0}+\beta\ket{1}) \otimes \ket{0}&\stackrel{
\textsc{clone}}{\longmapsto}(\alpha\ket{0}+\beta\ket{1}) \otimes (\alpha\ket{0}+\beta\ket{1})\\
\left(\begin{array}{c}
\alpha\\
0\\
\beta\\
0
\end{array}\right)
&\stackrel{\textsc{clone}}{\longmapsto} \left(\begin{array}{c}
\alpha^2\\
\alpha\beta\\
\alpha\beta\\
\beta^2
\end{array}\right),
\end{align*}
which cannot be linear (a more formal discussion can be found in
\cite{Wootters}). Cloning should be distinguished from
\emph{copying} however, as we now illustrate once more on qubit:

\begin{tabular}{c@{}c}
$\begin{array}{@{}rcl}
\ket{0} \otimes \ket{0}&\stackrel{\textsc{copy}}{\longmapsto}\ket{0}
\otimes \ket{0}\\
\ket{1} \otimes \ket{0}&\stackrel{\textsc{copy}}{\longmapsto}\ket{1}
\otimes \ket{1}\\
(\alpha\ket{0}\!+\!\beta\ket{1}) \otimes \ket{0}&\stackrel{\textsc{copy}}{\longmapsto}
\alpha\ket{0} \otimes \ket{0}+\beta\ket{1} \otimes \ket{1}
\end{array}$
&
$\begin{array}{rcl}
\left(\begin{array}{c}
\alpha\\
0\\
\beta\\
0
\end{array}\right)
&\stackrel{{\textsc{copy}}}{\longmapsto} \left(\begin{array}{c}
\alpha\\
0\\
0\\
\beta
\end{array}\right).
\end{array}$
\end{tabular}

Such an evolution is perfectly valid, and in the above case it may
be implemented as a single application of the quantum gate $CNOT$,
which is of course both linear and unitary.\\
In classical functional languages terms such as $\lambda
x.f(x,x)$, with $(\lambda x.f(x,x)\; t)
\stackrel{\beta}{\longrightarrow}f(t,t)$, are crucial for the
expressiveness. Recursion, for instance, relies upon such terms,
and is absolutely necessary for universality. When designing a
quantum functional language we therefore face a choice:
\begin{itemize}
\item Either we prevent terms such as $\lambda x.f(x,x)$ from being
applied to quantum states --- thereby ensuring that no quantum
cloning is allowed. But we must authorize their applications upon
``classical terms'' for expressiveness. As a consequence the
language must be able to keep track of quantum resources versus
classical resources. This is the approach followed by Van Tonder
\cite{Van Tonder1}.
\item Or we allow terms such as $\lambda x.f(x,x)$ from being
applied to quantum states --- only to be interpreted as a quantum
copy.  We may still want to keep track of quantum resources versus
classical resources, but not for the purpose of forbidding
cloning. This is the approach we take.
\end{itemize}
For now the latter option seems preferable, since it models the
\emph{dos} and \emph{don'ts} of the linearity requirement more
closely, whilst keeping the calculus to a minimum. Moreover
copying, as we now show, can be imposed over cloning by the
semantics of the calculus alone. Thus, the fact that classical states
can be cloned is proved and not postulated in our language.

\subsection{Substitution of de Bruijn indices}

The point of the previous discussion is that we can only duplicate
basis vectors. Informally
\def\sotimes{{\otimes}}
\begin{align*}
(\lambda x .(x \sotimes x))*\textbf{true} &\lr^*
\textbf{true} \otimes \textbf{true}\hspace{4ex}\textrm{is OK;}\\
(\lambda x .(x \sotimes x))*(\textbf{false}+\textbf{true}) &\lr^*
\big((\lambda x .(x \sotimes x))*\textbf{false}\big) + \big((\lambda
x.(x \sotimes x))*\textbf{true}\big)\\
&\lr^* \big(\textbf{true} \otimes \textbf{true}\big) +
\big(\textbf{false} \otimes \textbf{false}\big)\hspace{2.5ex}\textrm{is OK;}\\
(\lambda x .(x \sotimes x))*(\textbf{false}+\textbf{true}) &\lr^*
(\textbf{false}+\textbf{true}) \otimes
(\textbf{false}+\textbf{true})\hspace{3ex}\textrm{is not OK. }
\end{align*}
Again another way to grasp this idea is to realize that faced with
a term of the form $(\lambda x. t) * (u+v)$, one could either
start by proceeding to the substitution, or start by applying the
right-hand-side linearity of *, leading to two different results.
So that operations remain linear, we must favour the
right-hand-side linearity of * over substitution. The rules of
figure \ref{ruleslinoversub} accomplish exactly that.
\begin{figure}[!h]
\center{\fbox{\begin{minipage}[t]{11.3cm}\vspace{-2mm}
\caption{\label{ruleslinoversub}\small\textsc{Enforcing linearity
over substitution }}\vspace{-6mm}
\begin{align*}
&L(\mathbf{u}) * \mathbf{v} \lr\; (\mathbf{u} \textrm{ of } \mathbf{v})\\
&\mathbf{t} \textrm{ of } (r . \mathbf{v}) \lr\; r . (\mathbf{t}
\textrm{ of } \mathbf{v})\\
&\mathbf{t} \textrm{ of } (\mathbf{v} + \mathbf{w}) \lr\;
(\mathbf{t} \textrm{ of } \mathbf{v}) + (\mathbf{t} \textrm{ of }
\mathbf{w})\\
&\mathbf{t} \textrm{ of } \textbf{true} \lr\; \mathbf{t} \textrm{ bof } \textrm{subst}(\textbf{true})\\
&\mathbf{t} \textrm{ of } \textbf{false} \lr\; \mathbf{t} \textrm{ bof } \textrm{subst}(\textbf{false})\\
&\mathbf{t} \textrm{ of } \mathbf{0} \lr\; \mathbf{t} \textrm{ bof
} \textrm{subst}(\mathbf{0})\\
&\mathbf{t} \textrm{ of } (\mathbf{v} \otimes \mathbf{w}) \lr\\
&\left(\Big(\big((\mathbf{t} \textrm{ bof }  \Uparrow\!(\uparrow))
\textrm{ bof } \Uparrow\!(\uparrow)\big) \textrm{ bof }
\textrm{subst}\big(\textbf{var}(0) \otimes
\textbf{var}(S(0))\big)\Big) \textrm{ of }
\mathbf{v}\right)\textrm{ of } \mathbf{w}\\
&\mathbf{t} \textrm{ of } (\mathbf{v} \rhd \mathbf{w}) \lr\\
&\left(\Big(\big((\mathbf{t} \textrm{ bof }  \Uparrow\!(\uparrow))
\textrm{ bof } \Uparrow\!(\uparrow)\big) \textrm{ bof }
\textrm{subst}\big(\textbf{var}(0) \!\rhd\!
\textbf{var}(S(0))\big)\Big) \textrm{ of }
\mathbf{v}\right)\textrm{ of } \mathbf{w}
\end{align*}
\end{minipage}}}
\end{figure}
The three first rules are the most straightforward, they invoke
the linearity of the vector to be substituted. The three following
rules treat the base cases, when the vector to be substituted is
down to a basic state. The last two rules handle the more subtle
case of tensor states $\mathbf{u \otimes v}$ or $\mathbf{u\rhd v}$.
In a word the trick is to treat
\begin{align*}
 &\big(\lambda x.(\hdots x\hdots)\big)*(\mathbf{u \otimes
 v})\quad\textrm{as}\quad\Big(\lambda x.\big(\lambda y.(\hdots x \otimes y
 \hdots)\big)*\mathbf{u}\Big)*\mathbf{v}
\end{align*}
with $y$ a fresh variable, and then proceed recursively.\\
Once the vector to be substituted is a basic state, we can safely
proceed to the substitution using a calculus of explicit substitutions 
\cite{Abadi,Curien,Lescanne}. Here we have chosen to represent
avariables by their de Bruijn indices, i.e. each variable is
now an integer number corresponding to number of binders (``$L$'' or
``$\lambda$'' symbols) one must go through before reaching the
binding occurrence. For instance
\begin{equation*}
\lambda x.(\lambda y. (x \otimes y))\textrm{  is encoded as  }
L(L(\va(1) \otimes \va(0)))
\end{equation*}
since there is one $\lambda$ symbol lying between $x$ and the
binding occurrence of $x$. This variable numbering scheme is often
used for implementing functional languages. Notice how in this
scheme a variable may be denoted differently depending upon its
position in the term (i.e. depending upon how far it lies from its
binding occurrence). The rules of figure \ref{rulesbruijn}
implement this mechanism.
\begin{figure}[!h]
\center{\fbox{\begin{minipage}[t]{11.3cm}\vspace{-2mm}
\caption{\label{rulesbruijn}\small\textsc{Explicit substitution of
de Bruijn indices }}\vspace{-6mm}
\begin{align*}
(\mathbf{t} + \mathbf{u}) \textrm{ bof } \mathbf{s} &\lr
(\mathbf{t} \textrm{ bof } \mathbf{s}) + (\mathbf{u} \textrm{ bof
}\mathbf{s})\\
(r . \mathbf{u}) \textrm{ bof } \mathbf{s} &\lr (r \textrm{ bof }
\mathbf{s}) . (\mathbf{u} \textrm{ bof } \mathbf{s})\\
(\mathbf{t} \bullet \mathbf{u}) \textrm{ bof } \mathbf{s} &\lr
(\mathbf{t} \textrm{ bof } \mathbf{s}) \bullet (\mathbf{u}
\textrm{ bof } \mathbf{s})\\
(r + s) \textrm{ bof } \mathbf{s} &\lr (r \textrm{ bof } \mathbf{s}) + (s\textrm{ bof } \mathbf{s})\\
(r \times s) \textrm{ bof } \mathbf{s} &\lr (r \textrm{ bof } \mathbf{s}) \times (s \textrm{ bof } \mathbf{s})\\
r \textrm{ bof } \mathbf{s} &\lr r\\
(\mathbf{t} \otimes \mathbf{u}) \textrm{ bof } \mathbf{s} &\lr (\mathbf{t} \textrm{ bof } \mathbf{s}) \otimes (\mathbf{u} \textrm{ bof } \mathbf{s})\\
(\mathbf{t} \rhd \mathbf{u}) \textrm{ bof } \mathbf{s} &\lr
(\mathbf{t} \textrm{ bof } \mathbf{s}) \rhd (\mathbf{u} \textrm{
bof } \mathbf{s})\\
 (\mathbf{t} * \mathbf{u}) \textrm{ bof }
\mathbf{s} &\lr (\mathbf{t} \textrm{ bof } \mathbf{s})
* (\mathbf{u} \textrm{ bof } \mathbf{s})\\
L(\mathbf{t}) \textrm{ bof } \mathbf{s} &\lr L(\mathbf{t} \textrm{ bof } \Uparrow(\mathbf{s}))\\
\textbf{0} \textrm{ bof } \mathbf{s} &\lr \textbf{0}\\
\textbf{false} \textrm{ bof } \mathbf{s} &\lr \textbf{false}\\
\textbf{true} \textrm{ bof } \mathbf{s} &\lr \textbf{true}\\
\textbf{var}(0) \textrm{ bof } \textrm{subst}(\mathbf{v}) &\lr \mathbf{v}\\
\textbf{var}(S(p)) \textrm{ bof } \textrm{subst}(\mathbf{v}) &\lr \textbf{var}(p)\\
\textbf{var}(0) \textrm{ bof } \Uparrow(\mathbf{s}) &\lr \textbf{var}(0)\\
\textbf{var}(S(p) ) \textrm{ bof } \Uparrow(\mathbf{s}) &\lr
(\textbf{var}(p) \textrm{ bof } \mathbf{s}) \textrm{ bof } \uparrow\\ \textbf{var}(p)     \textrm{ bof } \uparrow &\lr \textbf{var}(S(p))
\end{align*}
\end{minipage}}}
\end{figure}

Thus there are two ways to use the $\lambda$-calculus to define a linear
map in a vectorial space of countable dimension over a countable
field. The first is to interprete the $\lambda$-terms as functions
mapping vectors to vectors and in this case we need extra constraints
to enforce linearity. The second is to interprete the $\lambda$-terms as
functions mapping base vectors to base vectors and to extend it to
the full space with extra computation rules. This solution is
advantageous as it requires no restriction on the $\lambda$-terms.

\begin{figure}[!ht]
\center{\begin{minipage}[t]{11.3cm}\vspace{-2mm}
\caption{Summary}

\begin{center}
\begin{tabular}{|c|c|}\hline 
Computational scalars           & {\em (rules Fig. \ref{rulesfloats},
\ref{rulesscals} and \ref{rulesvecsp})} \\
\hline 0                       & null scalar \\
\hline 1                       & unit scalar \\
\hline +                       & scalar sum \\
\hline $\times$                & scalar product \\

\hline
\hline Vectorial space   & {\em (rules Fig. \ref{rulesvecsp})} \\
\hline {\bf 0}                 & null vector\\
\hline {\bf false}, {\bf true} & base vectors \\
\hline +                       & vector sum \\
\hline .                       & product of a vector by a scalar \\
                                    
\hline
\hline Tensorial product & {\em (rules Fig. \ref{rulestensors})} \\
\hline $\otimes$                 & tensorial product\\

\hline
\hline Matching operators & {\em (rules
Fig. \ref{rulesmatchingbilinearity}, \ref{rulesmatchingtoscalar})} \\
\hline $\rhd$                   & matching construct\\
\hline *                        & function application \\
\hline $\bullet$                & scalar product\\

\hline \hline Lambda calculus & {\em (rules Fig. \ref{ruleslinoversub}, \ref{rulesbruijn})}\\
\hline $L$                      & lambda abstraction\\
\hline {\bf var}                   & variables\\
\hline of, bof, subst, $\uparrow$, $\Uparrow$ & explicit substitutions
constructs \\ 
\hline
\end{tabular}
\end{center}
\end{minipage}}
\end{figure}

Notice that our language allows a restricted form of higher-order
programming where a function $F$ defined by a $\lambda$-term can take
another function $g$ as its argument, provided the function $g$ is
expressed with the matching construct, but not if it is expressed as
a $\lambda$-term also. The extention of this language to full
higher-order programming is left for future work.

\section{An example}
\label{example}

The Deutch-Jozsa algorithm can be defined in our language as follows 
$$DJ = 
\lambda x~((Cross * H * H) * (x * ((Cross * H * H) * ({\bf false}
\otimes {\bf true}))))$$
where 
\begin{align*}
Cross  &= \lambda x~\lambda y~(\\
&({\bf false} \otimes {\bf false}) \rhd 
((x * {\bf false}) \otimes (y * {\bf false}))\\
&+ ({\bf false} \otimes {\bf true}) \rhd 
((x * {\bf false}) \otimes (y * {\bf true}))\\
&+ ({\bf true} \otimes {\bf false}) \rhd 
((x * {\bf true}) \otimes (y * {\bf false}))\\
&+ ({\bf true} \otimes {\bf true}) \rhd 
((x * {\bf true}) \otimes (y * {\bf true})))\\
\end{align*}
i.e. using de Bruijn indices 
$$DJ = 
L ((Cross * H * H) 
* ({\bf var}(0) * ((Cross * H * H) * 
({\bf false} \otimes {\bf true}))))$$
where 
\begin{align*}
Cross  &= L(L(\\
&({\bf false} \otimes {\bf false}) \rhd 
(({\bf var}(S(0)) * {\bf false}) \otimes ({\bf var}(0) * {\bf false}))\\
&+ ({\bf false} \otimes {\bf true}) \rhd 
(({\bf var}(S(0)) * {\bf false}) \otimes ({\bf var}(0) * {\bf true}))\\
&+ ({\bf true} \otimes {\bf false}) \rhd 
(({\bf var}(S(0)) * {\bf true}) \otimes ({\bf var}(0) * {\bf false}))\\
&+ ({\bf true} \otimes {\bf true}) \rhd 
(({\bf var}(S(0)) * {\bf true}) \otimes ({\bf var}(0) * {\bf true}))))\\
\end{align*}
Then it can be checked that the term $DJ * CNOT$ reduces
to ${\bf true} \otimes {\bf true}$ whose
first component is indeed the
exclusive disjunction of ${\bf not}({\bf true})$ and ${\bf not}({\bf
false})$.

\section{Discussion}
\label{sectiondiscussion}
\subsection{Unitarity}
In its simplest formulation quantum theory only allows unitary
evolutions, i.e. vectors evolve in time according to square
matrices $U$ verifying $U^{\dagger}U=\mathbb{I}$. In this
framework it is impossible to delete, say, a qubit:
\begin{align*}
\ket{0}&\stackrel{\textsc{erase}}{\longmapsto}\ket{0}\\
\ket{1}&\stackrel{\textsc{erase}}{\longmapsto}\ket{0}\\
(\alpha\ket{0}\!+\!\beta\ket{1})&\stackrel{\textsc{erase}}{\longmapsto}\ket{0},
\end{align*}
the evolution is not injective and therefore not unitary. Von
Neumann's projective measurements help us only partially: if the
vector is measured in the canonical basis, and when this
measurement yields outcome ``0'', then the qubit undergoes the above
exact dynamics. But this will only occur with probability
${|\alpha|}^2$.\\
The {\sc erase} operation is however perfectly physical, as one
can always ignore a qubit and focus upon another, taken to be in
state $\ket{0}$. Moreover the process needs not be probabilistic.
There are two well-established formulations of quantum theory
which cater for this possibility:
\begin{itemize}
\item The \emph{generalized measurement} formalism unifies quantum evolutions
and quantum measurements as one single object. Mathematically a
generalized measurement is given by a set of matrices $\{M_m\}$
verifying
\begin{equation}\label{trace preserving condition}
\sum_m M_m^{\dagger} M_m<\!\!<Id.
\end{equation}
A vector $v$ will then evolve in time according to the matrix
$M_m$ with probability $p_m=|M_m v|^2$ --- in which case we shall
say that outcome ``$m$'' has occurred. As an example the following
generalized measurement performs the \textsc{erase} operation:
\begin{equation*}
\left\{\left(\begin{array}{cc} 1  &0\\
0  &0 \end{array}\right),
\left(\begin{array}{cc} 0  &1\\
0  &0
\end{array}\right)\right\}.
\end{equation*}
For a more detailed presentation of these concepts the reader is
referred to \cite{Nielsen}, page 84.
\item The \emph{density matrix} formalism represents quantum
states as positive matrices instead of vectors. These evolve in
time according to Completely Positive-preserving maps, i.e.
operations of the form
\begin{equation*}
\rho\longmapsto\sum_m M_m \rho M_m^{\dagger} \quad \textrm{with
probability} \quad p_m=\trace(M_m \rho M_m^{\dagger})
\end{equation*}
and verifying Equation (\ref{trace preserving condition}). This
framework is traditional when dealing with open quantum systems,
and therefore well appropriate as one discards a qubit.
\end{itemize}
In classical functional languages terms such as $\lambda
y.(\lambda x.x)$, with $(\lambda y.(\lambda x.x) \mathbf{t})
\stackrel{\beta}{\longrightarrow}\lambda x.x$, are commonly used.
Boolean values and branching, for instance, are encoded in such
manners. Although convenient, non-injective functions are not
absolutely necessary, for reversible computation can be both
universal and efficient \cite{Bennett}. Whether a quantum
functional language should allow erasure or not must therefore
depend upon which of the three above mentioned formulation of
quantum theory gets chosen.

If we adopt the simplest formulation of quantum theory, our
language must be restricted to operation which are a not only
linear but also unitary. At first sight this seems feasible by
imposing the standard $U^{\dagger}U=\mathbb{I}$ condition upon the
matching constructs of section \ref{sectionmatching}, and the
\emph{relevance} condition upon the $\lambda$-terms of
\ref{sectionlambda} (i.e. for all term $\lambda x. t$ the variable
$x$ must occur at least once in $t$). This remains a
subject for future work.

\subsection{Linearity?}
Linear logic appears in \cite{Girard1} as a mean to express and
prove properties of dynamical systems where the \emph{consumption}
of resources is important. The standard example (price updated) is
$A\equiv$``I have $6\europ$'', $B\equiv$``I have a paquet of
Gauloises'', and the statement $A\multimap B$ to express the
possibility of using up $A$ to obtain $B$. With $\otimes$ now
expressing a conjunction, it is clear one cannot have $A\multimap
(B \otimes B)$, since this would mean buying two paquets for the
price of one. Neither can we have $A \otimes A\multimap B$: we must
get something for our money --- at worse the feeling of getting
cheated. Therefore the rules governing symbols $\multimap$,
$\otimes$ differ from those of classical logic for $\Rightarrow$,
$\wedge$. Unless there is an abundance of resources (denoted by
the exclamation mark ``!''), in which is case they coincide again:
$!A\multimap (B \otimes B \otimes \ldots)$. Whilst considering this
point the father of Linear logic has the following thought
\cite{Girard2}: ``\emph{Classical logic appears to be the logic of
macro-actions, as opposed to linear logic which would be a logic
of micro-actions. The unusual character of linear logic may
therefore be considered similar to the strange character of
micro-mechanics, i.e. quantum mechanics.}''.

Specifications expressed in linear logics can be seen as types for
programs expressed in linear $\lambda$-calculus. In the linear
$\lambda$-calculus one distinguishes \emph{linear resources},
which may not be copied nor discarded, from \emph{nonlinear
resources}, which are denoted by the exclamation mark ``!'' and
whose fate is not subjected to particular restrictions. Van
Tonder's quantum $\lambda$-calculus ($\lambda_q$) is founded upon
these ideas. As we have mentioned in \ref{subsectioncloning}, he
uses this well-established framework in order to distinguish
quantum resources (treated as linear) from classical resources
(treated as nonlinear):
\begin{align*}
(\lambda_q)\qquad &t::=x\,|\,\lambda x.t\,|\,(t\;\;t)\,|\,c\,|\,!t\,|\,\lambda !x.t\\
&c::=0\,|\,1\,|\,H\,|\,CNOT\,|\,P\\
&\textrm{(together with the well-formedness rules of the classical linear}\\
&\lambda\textrm{ calculus)}\\
&~\\[-1ex]
(\mathcal{R}_q)\qquad &(\lambda x.t\;\;s)\stackrel{\beta}{\longrightarrow} t[s/x]\\
&H\;\;0\longrightarrow 0+1\\
&H\;\;1\longrightarrow 0-1\\
&\vdots
\end{align*}
Here the well-formedness conditions of the classical linear
calculus (which prevents linear terms from being discarded),
together with $!$-suspension (which stops quantum terms from being
treated as nonlinear) maintain unitarity throughout the
reductions.

The connection between the linear $\lambda$-calculus and quantum
functional languages is striking. It comes at a price however: the
$\lambda_q$-calculus remains heterogeneous, i.e. a juxtaposition
of quantum resources (linear resources) and classical resources
(nonlinear resources). In some sense it is twice linear, both in
the sense of linear $\lambda$-calculus and linear algebra, and
thus in some sense overrestricted. In particular it forbids both
the cloning of quantum data (which needs be done) and the copying
of quantum data (which needs not be done). As a consequence its
control flow remains inherently based upon classical resources.

The \emph{linear-algebraic $\lambda$-calculus} constructed in this
paper is homogeneous, i.e. it does not draw a line between quantum
resources and classical resources (the latter are merely thought
of as basis states of the former). Moreover it exhibits only one
notion of linearity, which is that of linear algebra. Thus cloning
remains disallowed (just by the semantics) but not copy. Control
flow is still provided as a consequence. These results seem to
open the way to a linear algebraic interpretation of linear logic,
in the spirit of Girard's Geometry of interaction, although much
work remains ahead in order to strengthen this connection.

\end{document}